\documentclass[twocolumn,aps,prb,showpacs,superscriptaddress,floats,amsmath,amssymb,floatfix]{revtex4}

\usepackage{color}
\usepackage{amsmath}
\usepackage{graphicx}
\usepackage{graphics}
\usepackage[pdf]{pstricks}
\usepackage{pdftexcmds}
\usepackage{amssymb}
\usepackage{esint}
\usepackage{comment}
\usepackage{physics}
\usepackage{bm}
\usepackage{lipsum}    
\usepackage[utf8]{inputenc}

\def \bea {\begin{eqnarray}}
\def \eea {\end{eqnarray}}

\allowdisplaybreaks

\begin{document}

\title{Machine learning techniques  to construct  detailed phase diagrams for  skyrmion systems  }

\author{F. A. G\'omez Albarrac\'in}
\email[corresponding author: ]{albarrac@fisica.unlp.edu.ar}
\affiliation{Instituto de Física de Líquidos y Sistemas Biológicos (IFLYSIB), UNLP-CONICET, Facultad de Ciencias Exactas, La Plata, Argentina}
\affiliation{Departamento de Física, Facultad de Ciencias Exactas, Universidad Nacional de La Plata, La Plata, Argentina}
\affiliation{Departamento de Ciencias Básicas, Facultad de Ingeniería, Universidad Nacional de La Plata, La Plata, Argentina}
\author{H. D. Rosales}
\affiliation{Instituto de Física de Líquidos y Sistemas Biológicos (IFLYSIB), UNLP-CONICET, Facultad de Ciencias Exactas, La Plata, Argentina}
\affiliation{Departamento de Física, Facultad de Ciencias Exactas, Universidad Nacional de La Plata, La Plata, Argentina}
\affiliation{Departamento de Ciencias Básicas, Facultad de Ingeniería, Universidad Nacional de La Plata, La Plata, Argentina}

\date{\today}

\begin{abstract}

Recently, there has been an increased interest in the application of machine learning (ML) techniques to a variety of problems in condensed matter physics. In this regard, of particular significance is  the characterization of simple and complex phases of matter. Here, we use a ML approach to construct the full phase diagram of a well known spin model combining ferromagnetic exchange and Dzyaloshinskii-Moriya (DM) interactions where topological phases emerge. At low temperatures, the system is tuned from a spiral phase to a skyrmion crystal by a magnetic field. However, thermal fluctuations induce two types of intermediate phases, bimerons and skyrmion gas, which are not as easily determined as spirals or skyrmion crystals.   We resort to large scale Monte Carlo simulations to obtain low temperature spin configurations, and train a convolutional neural network (CNN), taking only snapshots at specific values of the DM couplings, to classify between the different phases, focusing on the intermediate and intricate topological textures. We then apply the CNN to higher temperature configurations and to other DM values, to construct a detailed magnetic field-temperature phase diagram, achieving outstanding results.   We discuss the importance of including the disordered paramagnetic phases in order to get the phase boundaries, and finally, we compare our approach with other ML algorithms.

\end{abstract}

\maketitle


\section{Introduction}

In the last five years, Machine Learning (ML) techniques have provided a new perspective on the study of a great variety of physical phenomena  in condensed matter physics, from representation of quantum states \cite{Troyer} to discovering phase transitions \cite{Rem,Wang,Nieuwenburg} and identifying conventional phases of matter \cite{Melko}. The ability of ML  to identify and classify huge data sets, including images, provides a powerful tool to analyse the state space of condensed-matter systems.

These approaches have been successfully applied to a variety of complex topological spin systems and models. For example, autoencoders have been used to extract models from neutron scattering data in spin ice systems \cite{MLGrigera1,MLGrigera2}. Support vector machines with a tensorial kernel have been used to explore tensor order parameters and hidden order in non-trivial frustrated models such as the XXZ model on the pyrochlore lattice \cite{PolletPiro}, the classical kagome antiferromagnet \cite{PolletKagome}, and Kitaev models and materials \cite{PolletKitaev1,PolletKitaev2,PolletKitaev3}.

Among   exotic phases, magnetic skyrmions are  undoubtedly at the heart of a large body of work \cite{Review2017,BeyondSky,Review2021}.  Skyrmions are swirling magnetic textures characterized by a topological invariant, the topological charge $Q=\frac{1}{4\pi}\int d^2r 
\vec{S}\cdot(\partial_x\vec{S}\times\partial_y\vec{S})$ (where $\vec{S}$ is
the unit vector of the local magnetization), which provides them of great
stability against perturbations. Therefore,
magnetic skyrmions have been intensively studied due to their potential
in future magnetic data storage and spintronics applications.   In fact, neural networks have been implemented in the last years to distinguish skyrmion phases. On one hand, a few years ago, it was shown that a single layer neural network can succesfully classify standard configurations: spiral, ferromagnetic and skyrmion crystal \cite{MLSky}, and  a similar classification task was achieved with convolutional neural networks (CNNs) to construct  low temperature phase diagrams for models including anisotropy terms \cite{MLSkyJMMM}.  Recurrent neural networks were used to classify skyrmion dynamic processes \cite{IakovlevRR}. On the other hand, CNNs were used to predict features such as chirality in these type of systems \cite{MLSky2},  even in confined geometries \cite{MLSkyConf}, and to extract information on the interactions from data  images \cite{MLSkyDM}. Moreover, taking input data from videos, these type of techniques  were used to classify the dynamical skyrmion phases, and to predict phase boundaries \cite{MLSkyVideo}.  Most studies in this field have only focused on the characterization of typical configurations and not so much attention has been paid to the intermediate and less conventional phases emerging from thermal fluctuations.  

In this regard, the aim of this investigation is to explore the capacity of  machine learning algorithms to identify and classify all the magnetic phases in skyrmion systems. Our main goal is to apply these techniques to construct a complete magnetic field-temperature (B-T) phase diagram for a well-known model where a skyrmion crystal is stabilized at low temperatures, the ferromagnetic square lattice with Dzyaloshinskii-Moriya (DM) interactions under an external magnetic field. Magnetic configurations for this study were collected using  large scale Monte Carlo simulations. Then, we resort to a convolutional neural network (CNN) to classify the lowest temperature phases, that include bimeron   and skyrmion gas labels.  The term ``bimeron'' here is used to refer to elongated skyrmions or broken spirals \cite{Ezawa2011,RSilva,Leonov}, to differentiate from the use of the term which refers to two merons \cite{GobelBim}.

As a first step, we train and validate the network for three specific values of the DM interaction . We then apply the trained model to other DM values and to higher temperature configurations, and build the complete B-T phase diagram, which we compare with results from simulations, finding a remarkable agreement. In particular, our  CNN-based approach shows that all types of topological phases considered could be distinguished and classified.  Given that CNNs are techniques ideal for treating image data, this opens the door to use a similar approach for analysis of the experimental images obtained with spin-polarised scanning tunnelling microscopy techniques, where these type of intermediate phases are usually found. \cite{Yu2010,Bergman2014,Gao2019,Yasui2020}.

The manuscript is organised in the following way. In Sec. II we describe the model, discuss the emergent topological phases and show the particular features of the bimeron and skyrmion gas configurations. The ML approach and analysis are described in Sec. III, where we also explore  other simpler ML methods. Conclusions and future perspectives are presented in Sec. V.

\section{Skyrmion model}

\begin{figure}[h!]
\includegraphics[width=1.\columnwidth]{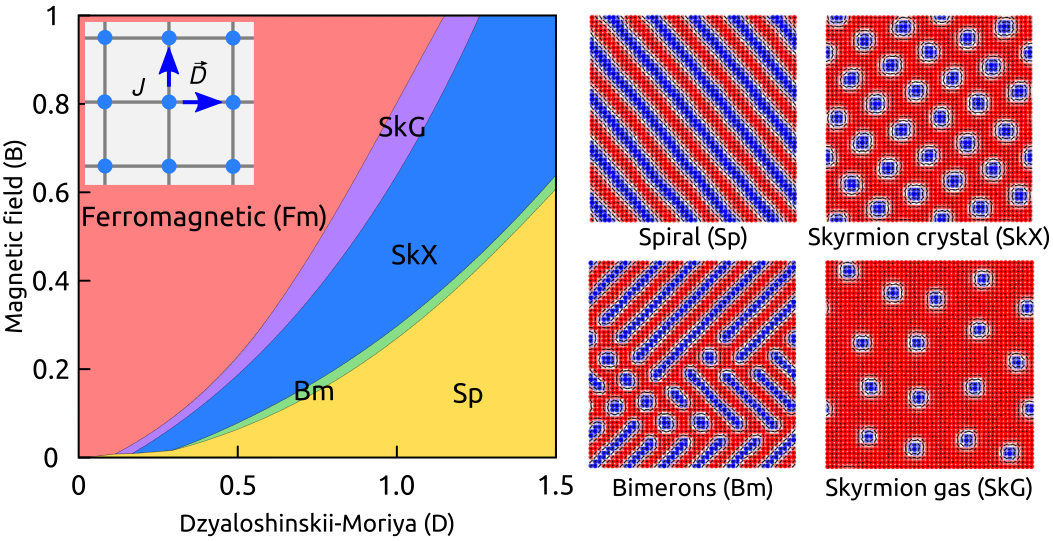} 
\caption{Left: low temperature phase diagram for the ferromagnetic model in a square lattice with DM interactions under a magnetic field $B$.  In the inset, (blue) arrows indicate the direction of the DM vectors, along the bonds of the square lattice. Right: typical real space configurations for the four different non-polarized phases: spiral (Sp), bimeron (Bm), skyrmion crystal (SkX), skyrmion gas (SkG).}
\label{fig:phases}
\end{figure}

 Skyrmions have been at the heart of a large body of work since the inderect experimental observation of skyrmion crystals through neutron diffraction in MnSi \cite{MnSi} and  direct observation in thin films  \cite{Yu2010}. Research has also expanded to other topological textures \cite{BeyondSky,Review2021}, such as antiferromagnetic skyrmions\cite{Rosales2015,AFSky1,AFSky2,Gao2020}, magnetic bubbles, merons and antimerons \cite{Yu2021}. Here, we focus in a well known skyrmion model, which was proposed to compare with one of the first bidimensional experimental phase diagrams \cite{Yu2010}, and has also been used to explore machine learning techniques for skyrmion phases \cite{MLSky,MLSky2}. We take the ferromagnetic exchange model with in plane Dzyaloshinskii-Moriya interactions (DM) under a magnetic field for classical Heisenberg spins in the square lattice, given by:
 
\begin{equation}
\mathcal{H}=-J\sum_{\langle i,j \rangle} \vec{S}_i\cdot\vec{S}_j + \sum_{\langle i,j \rangle} \vec{D}\cdot (\vec{S}_i\times\vec{S}_j) - \vec{B}\sum_i \vec{S}_i
\end{equation}

\noindent where $\vec{S}_i$ are Heisenberg spins at site $i$ with fixed norm 1, $J$ is the ferromagnetic exchange coupling, $\vec{D}$ the in-plane DM interaction  along the bonds of the square lattice (see inset in Fig.~\ref{fig:phases} and $\vec{B}=B\breve{z}$ the external magnetic field, perpendicular to the lattice plane. We will take $J=1$ thorough the rest of this work.

We briefly review here the well known behavior of this model with magnetic field and temperature, to motivate our study. We present in  Fig.~\ref{fig:phases}  the low temperature phase diagram ($T\ll J$) as a function of the magnetic field $B$ and the DM strength $D$. Regarding the magnetization process, the magnetic field takes the system from a spiral phase (Sp)  induced  at zero field for a small DM interaction \cite{Bogdanov1994} to a skyrmion crystal (SkX), and then to a ferromagnetic (Fm) state at higher $B$. There are two types of intermediate phases, which we will discuss further below: what we refer to here as a ``bimeron'' phase (Bm), which consists of a mixture of ``broken spirals'' or ``elongated skyrmions'' and skyrmions, and  a second ``skyrmion gas'' phase (SkG), where the skyrmions do no longer form a crystal, and are distributed in a ferromagnetic background.  These intermediate phases are enhanced by temperature, and should disappear in the zero temperature limit \cite{Ezawa2011}.  A detailed phase diagram for a similar system at larger system sizes may be found in Ref.[\onlinecite{IakovlevBim}]. We show typical real space configurations for different phases at the right of Fig.~\ref{fig:phases}.

\begin{figure}[h!]
\includegraphics[width=0.95\columnwidth]{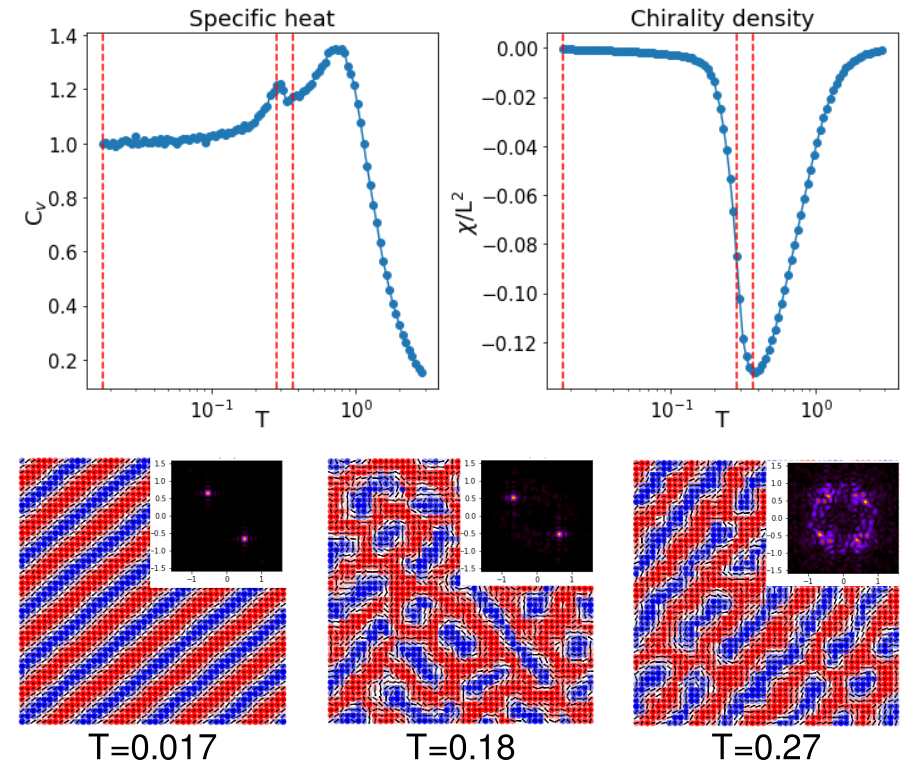} 
\caption{Monte Carlo simulations for $D=1$, $B=0.2$. Top panel: specific heat, magnetization and chirality as a function of temperature. Bottom panel: three snapshots and their corresponding $S_q$ for three different temperatures, indicated as vertical lines in the plots from the top panel. At the lowest simulated temperature, a spiral phase is stabilized, but an intermediate bimeron phase is clearly seen at higher $T$. }
\label{fig:BimT}
\end{figure}

\begin{figure}[h!]
\includegraphics[width=0.95\columnwidth]{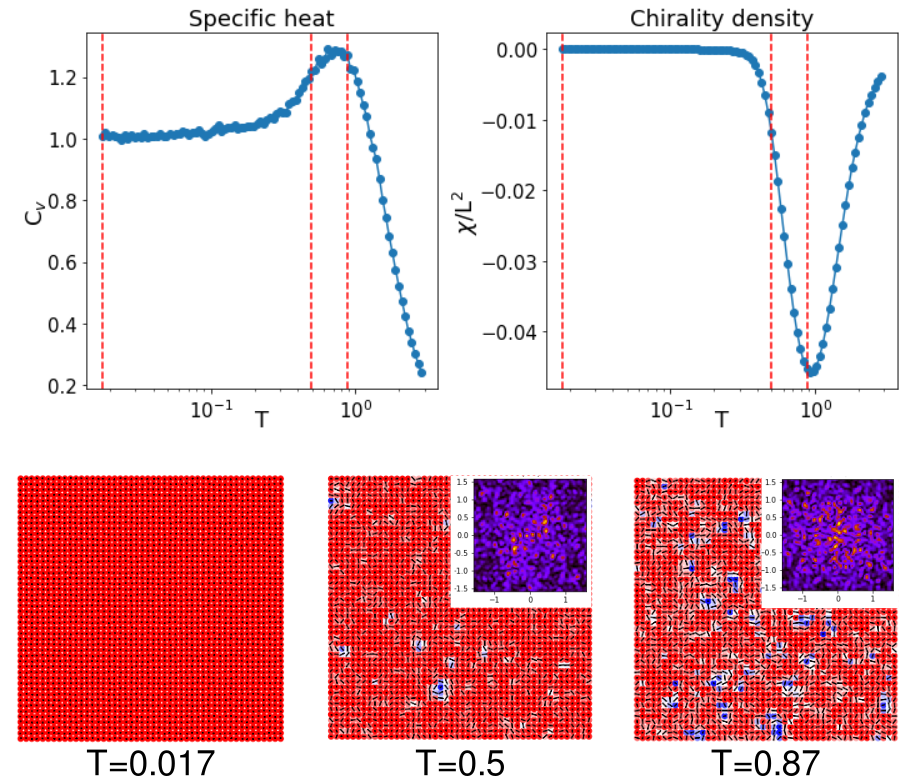} 
\caption{Monte Carlo simulations for $D=1$, $B=0.9$. Top panel: specific heat, magnetization and chirality as a function of temperature. Bottom panel: three snapshots and their corresponding $S_q$ for three different temperatures, indicated as vertical lines in the plots from the top panel. At the lowest simulated temperature, the system is in the ferromagnetic phase, but an intermediate skyrmion gas phase is clearly seen at higher $T$.}
\label{fig:SkGT}
\end{figure}

The spiral, skyrmion crystal and ferromagnetic phases can easily be distinguished resorting to variables such as the scalar chirality $\chi$ and the structure factor $S_{\mathbf{q}}$. Through all this work, $S_{\mathbf{q}}$  refers to the structure factor calculated with the components of the spins perpendicular to the external magnetic field, $S_{\mathbf{q}}=\frac{1}{N} \langle |\sum_{i} S_{i}^{x} \ e^{i {\bf q}\cdot {\bf r}_i} |^2 + |\sum_{i} S_{i}^{y} \ e^{i {\bf q} \cdot {\bf r}_i} |^2\rangle$, where $N$ is the total number of spins. The scalar chirality, defined as $\chi=\sum_{i,j,k}\vec{S}_i\cdot\left(\vec{S}_j\times\vec{S}_k\right)$, is quantity in  a lattice that is related to the calculation of the topological charge $Q$, which is $Q=-1$ for each skyrmion. Therefore, in the SkX phase $\chi$ is non zero and it is related to the number of skyrmions, whereas   the spiral and the ferromagnetic phase are not chiral, with $Q=0$. As for $S_{\mathbf{q}}$, a SkX is characterized with six Bragg peaks in reciprocal space, where the three inequivalent $\mathbf{q}^{*}$ vectors satisfy $\sum_i\mathbf{q}_i=0$ (triple-q phase).  In a spiral phase, for this model, there would be two inconmensurate Bragg peaks  (single-q). However, things are not so simple if we wish to take into account the intermediate phases. On one hand, the system does not present a clear transition between phases with temperature. On the other hand, specially at higher temperatures, these configurations do not have a characteristic structure factor. We illustrate this showing the results of Monte Carlo simulations for $D=1$ in Fig.~\ref{fig:BimT} ($B=0.2$) and Fig.~\ref{fig:SkGT} ($B=0.9$). In both cases, we plot the resulting specific heat, and chirality as a function of temperature, and show three different snapshots and their respective $S_q$ at different temperatures. In  Fig.~\ref{fig:BimT}, at $B=0.2$, we see that at low temperature the system is in a Sp phase, with the typical single-q $S_q$. At higher $T$, there is non-zero chirality and the snapshots show a Bm phase, which goes from  single-q to a less defined $S_q$. A similar behavior is seen at higher fields ($B=0.9$) in Fig.~\ref{fig:SkGT}, but comparing here a ferromagnetic phase at low temperature and an intermediate skyrmion gas phase. In the higher temperature snapshot, which corresponds to the highest value of the chirality, we can also see that although the system has a net chirality, thermal fluctuations break the skyrmion and less-defined chiral structures are seen. 

Following the discussion above, the identification of the intermediate phases and the  construction of a complete detailed $B-T$ phase diagram requieres a combination of resources, including calculation of the scalar chirality, inspection of the structure factor, and inspection of the real space configurations. In this work, our goal is to use machine learning techniques to assist in the detection of the intermediate ``non-standard'' phases and in the construction of this phase diagram, avoiding a snapshot-by-snapshot inspection. In the next section we describe our machine learning approach, and present our results using a Convolutional Neural Network.

\section{Machine Learning approach}

\begin{figure*}[th!]
\includegraphics[width=0.8\textwidth]{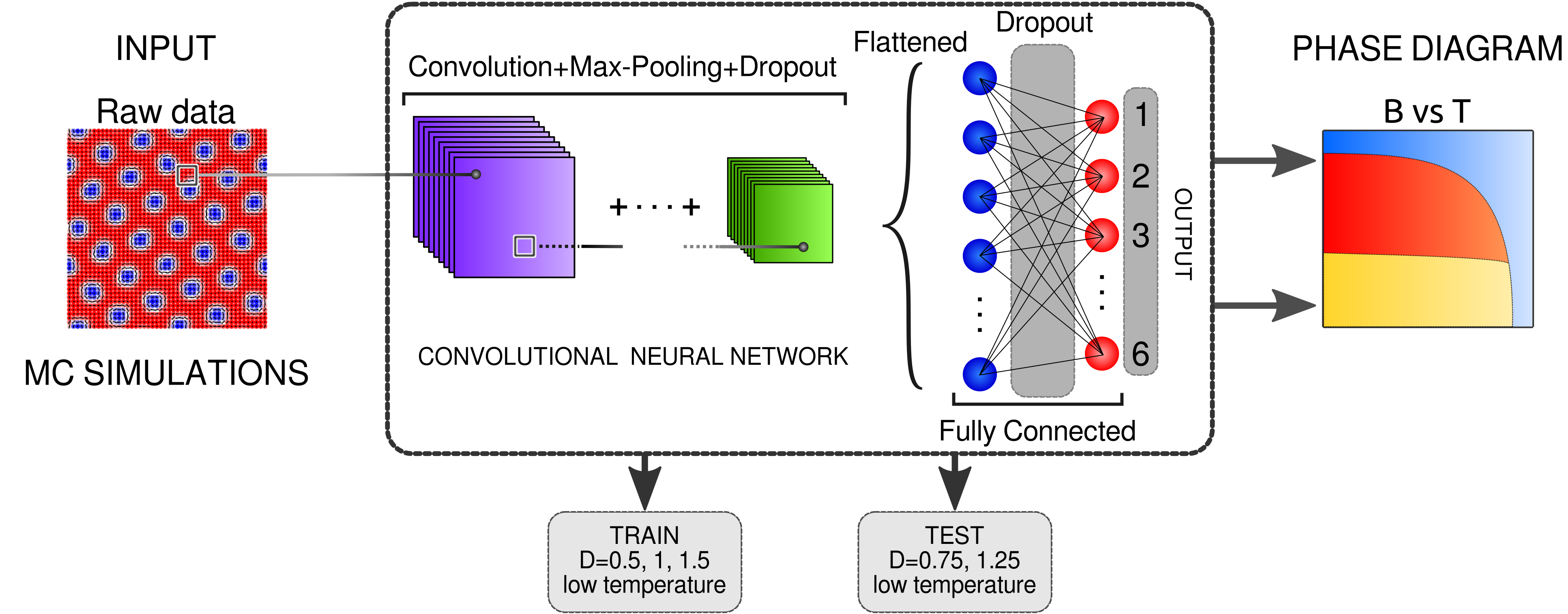} 
\caption{ Steps from the machine learning approach used in this work.  The dataset consists of $S^z$ components from spin configurations obtained from Monte Carlo simulations. A Convolutional Neural Network classification model was constructed, using the lowest temperature configurations at three values of the DM coupling ($D=0.5,1,1.5$) and a subset of high temperature data for training and validation, separating in six different phases: spirals, bimerons, skyrmion crystal, skyrmion gas, ferromagnetic and paramagnetic. A set of lowest temperature samples for different DM values ($D=0.75,1.25$) was used for testing. The resulting CNN classification model was applied for configurations in a wide range of temperature and magnetic field, to construct a complete $(B,T)$ phase diagram. Details of the CNN architecture are shown in the Appendix.   }
\label{fig:MLsteps}
\end{figure*}

Here we describe how we are going to use machine learning techniques to construct a complete $B-T$ phase diagram. We present the steps in Fig.~\ref{fig:MLsteps}.

First, in order to train and validate the ML algorithms we produce snapshots for the five different low temperature phases (Bm, Fm, SkG, SkX, Sp) using Monte Carlo simulations. We use the model Hamiltonian presented in Eq.~(1), but we only take three values of the DM interaction, $D=0.5,1.0,1.5$, for training and validation. Since we are taking low temperature configurations, from Fig.~ \ref{fig:phases} it can be clearly seen that the dataset will be very unbalanced, since the intermediate Bm and SkG phases are only stable in a very narrow range of magnetic field. Therefore, we run more simulations in these magnetic field ranges, in order to increase the number of Bm and SkG snapshots. The Monte Carlo simulations are done with the Metropolis algorithm combined with overrelaxation, for a $48\times 48$ square lattice, taking $10^5$ MC steps for thermalization and twice as much to take measurements. Each simulation is done fixing a value of $B$, using the annealing technique to take the system from $T=3$ to $T=0.017$. For each parameter set, $\sim 30$ copies with independent seeds were done. 

Secondly, we train a Convolutional Neural Network (CNN) for classification using the low $T$ dataset, where we have labeled the snapshots (this is an instance of \textit{supervised learning}).  As in previous works \cite{MLSky}, we only use as input the projection of the spins along the magnetic field, $S^z$.  This has the clear advantage of reducing the dataset by a factor of 3. Nonetheless, inclusion of the in-plane components of the spins would be needed to distinguish between Bloch and Néel skyrmions, and also antiskyrmions, which is not the case in this work.  We do not use any other variable ($D$, $B$, specific heat, chirality, etc.), to make sure that resulting model does not depend on parameters, but only on the configurations, and it can thus be potentially used for snapshots resulting from different Hamiltonians. Since we aim to construct a $B-T$ phase diagram, we also take snaphots at the highest temperatures, i. e. in the paramagnetic phase (Pm), at high ($B=0.9$) and zero magnetic field, where the system will not stabilize at low temperatures in a SkX phase. We will later discuss the effect of including these snapshots.

Thirdly, we apply the trained network to a test set, which we construct with low $T$ snapshots obtained for other values of the DM interaction, $D=0.75,1.25$, not used in training. The value of $D$ controls the size of the skyrmions and the spirals. Therefore, our idea here is to see whether the network can correctly identify topological structures in a parameter region where it has not been trained. This would be extremely useful, since it would mean that it is only necessary to train a technique in a small region of parameter space, and then it may be applied to any other values.

Finally, to construct the complete $B-T$ phase diagram, we apply the trained CNN to classify snapshots for all temperatures and magnetic fields. We do this for different values of $D$, and we compare the resulting phase diagrams with the chirality density phase diagram obtained from MC simulations. 

Below, we present a table detailing the dataset used in this work. 

\begin{table}[h!]
\centering
\begin{tabular}{|c|c|c|}
\hline 
Phases/Classes & $D=0.5,1.0,1.5$ & $D=0.75,1.25$ \\
\hline 
Bm & 208 & 134 \\
Fm & 330 & 10 \\
Pm & 600 & 60 \\
SkG & 240 & 90 \\
SkX & 260 & 308 \\
Sp & 360 & 360 \\
\hline 
Total & 1998 & 962 \\
\hline
\end{tabular}
\caption{Data set. Snapshots at $D=0.5,1.0,1.5$ were used for training and validation, and $D=0.75,1.25$ were used as the test set.  Bm: bimerons, Fm: ferromagnetic, Pm: paramagnetic, SkG: skyrmion gas,   SkX: skyrmion crystal, Sp: spiral }
\label{table:dataset}
\end{table}

\subsection{Convolutional Neural Network results}

Convolutional neural networks (CNN) have been used with great success in pattern and image recognition \cite{Geron}.  Details of the architecture of the CNN used in this work are given in the appendix.  The training dataset was split in 80$\%$ for training and 20$\%$ for validation, and the $S^z$ spin component was rescaled from $[-1,1]$ to $[0,1]$. The output of the CNN is a probability for each class (or phase), and we define the class of a given snapshot as the one that has the highest probability.


As a first step, we apply the CNN to the $D=0.5, 1.0, 1.5$ dataset, including all five low temperature phases and the paramagnetic phase. We obtain an accuracy of $99.9\%$ in the training set, and $97.5\%$ in the validation set. Then, we use this trained CNN to predict the phases for our test set, $D=0.75,1.25$, and obtain a slightly lower, but still satisfactory, accuracy of $91\%$. To analyze if there are specific phases where the CNN is missclassifying the snapshots, we calculate the confusion matrices (CM) for the training, validation and test set, which are to be read as follows: the i-th row indicates de true label for the i-th class, and the j-th row, the predicted label for the j-th class.  Therefore, at a quick glance, if the off-diagonal elements of the CM matrices are significantly lower than the diagonal ones, it would mean that there are few missclassified snapshots in that dataset.  We present the confusion matrices for the three sets below, where the classes (phases) are ordered alphabetically: (Bm, Fm, Pm, SkG, SkX, Sp):

\begin{equation}
CM_{train}=
\begin{pmatrix}
163 & 0 & 0 & 0 & 0 & 7\\
0 & 262 & 0 & 0 & 0 & 0 \\
0 & 0 & 473 & 0 & 0 & 0 \\
0& 5 & 0 & 187 & 0 & 0 \\
6 & 0 & 0 & 0 & 214 & 0\\
0 & 0 & 0 & 0 & 0 & 281 \\
\end{pmatrix}
\end{equation}

\begin{equation}
CM_{val}=
\begin{pmatrix}
36 & 0 & 0 & 0 & 1 & 1\\
0 & 68 & 0 & 0 & 0 & 0 \\
0 & 0 & 127 & 0 & 0 & 0 \\
0& 1 & 0 & 47 & 0 & 0 \\
1 & 0 & 0 & 0 & 39 & 0\\
0 & 0 & 0 & 0 & 0 & 79 \\
\end{pmatrix}
\end{equation}

\begin{equation}
CM_{test}=
\begin{pmatrix}
101 & 0 & 0 & 0 & 11 & 22\\
0 & 10 & 0 & 0 & 0 & 0 \\
0 & 0 & 60 & 0 & 0 & 0 \\
0& 3 & 0 & 87 & 0 & 0 \\
11 & 0 & 0 & 30 & 267 & 0\\
9 & 0 & 0 & 0 & 0 & 351 \\
\end{pmatrix}
\end{equation}

\begin{figure}[h!]
\includegraphics[width=0.95\columnwidth]{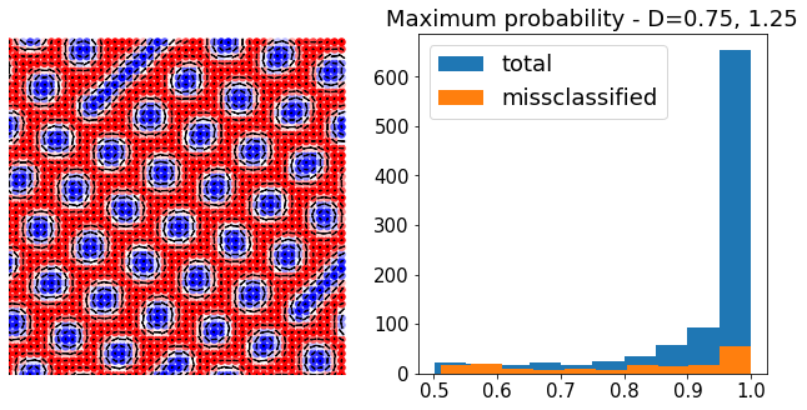}
\caption{\label{fig:miss} Left: Example of a missclassifed snapshot for the test dataset. The CNN assigns a $55\%$ probability to the SkX phase, and $44\%$ to the Bm phase. Right: Histograms of the maximum probability for the complete test set (blue) and for the missclassified configurations in the test set (orange).  It can be seen that there is a larger portion of missclassified snapshots with lower highest probabilities, and that the highest probabilities correspond to correctly classified configurations. }
\end{figure}

  Inspecting the CMs, we can see that, not surprisingly, the CNN has more trouble classifying the intermediate phases, Bm and SkG  (first and fourth rows) . Notice that the test dataset contains less ferromagnetic and paramagnetic phases (table \ref{table:dataset}), where the network has more accuracy: from the CMs, reading rows 2 and 3, it can be seen that all Fm and Pm snapshots where correctly classified.  To further analyze the missclassified data, we first present as an example,  a missclassified snapshot from the test set in Fig.~\ref{fig:miss}, left. This is a Bm phase, where there are some bimerons among skyrmions arranged in a lattice. If we look at the probabilities calculated for this configuration by the CNN, we see that it assigns it as a SkX with a probability of $55\%$, but the Bm class, which is the correct class, has the second highest probability, $44\%$.  Then, there is not a large difference between the two highest probabilities assigned by the CNN, i.e. the highest probability assigned is not significantly high.   We then plot in Fig.~\ref{fig:miss}, left, an histogram of the highest probability assigned to the chosen phase to the whole test set, and compare it with the same histogram obtained only for missclassified snapshots. We can clearly see that in the missclassified subset the CNN assigns much lower probabilities  than to the correctly classified ones, since the peak for the probabilities higher than 0.9 is not seen in the histogram of the missclassified data.  Comparing the counts from the histogram, it may also be seen that there are significantly less missclassified snapshots, suggesting a high accuracy.

\begin{figure}[h!]
\includegraphics[width=1.0\columnwidth]{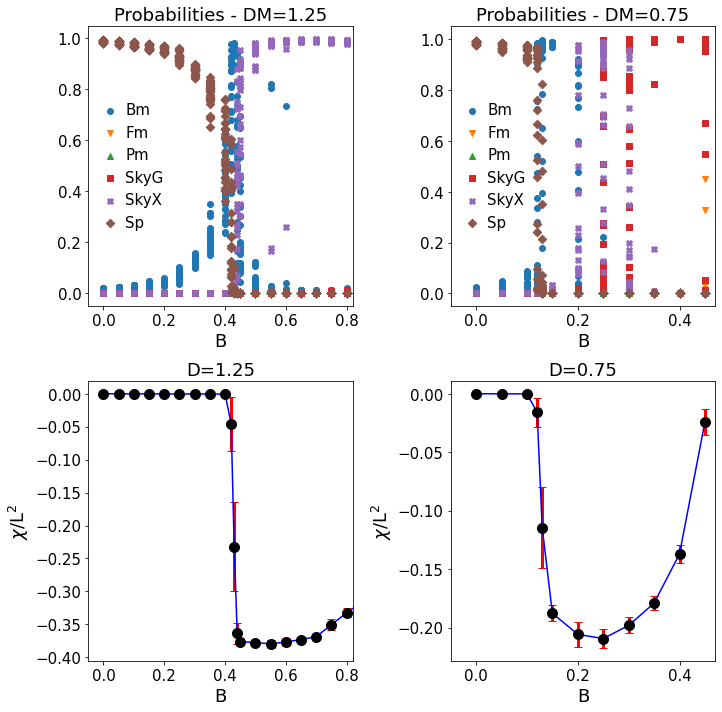}
\caption{\label{fig:probstest} Top: Probabilities calculated by the trained CNN as a function of $B$ for the test set values of the DM interaction, $D=1.25$ (left) and $D=0.75$ (right) Bottom: Chirality density obtained from MC simulations as a function of magnetic field, averaged over 30 independent copies. Errorbars indicate the standard deviation.}
\end{figure}

\begin{figure*}[t!]
\includegraphics[width=0.95\textwidth]{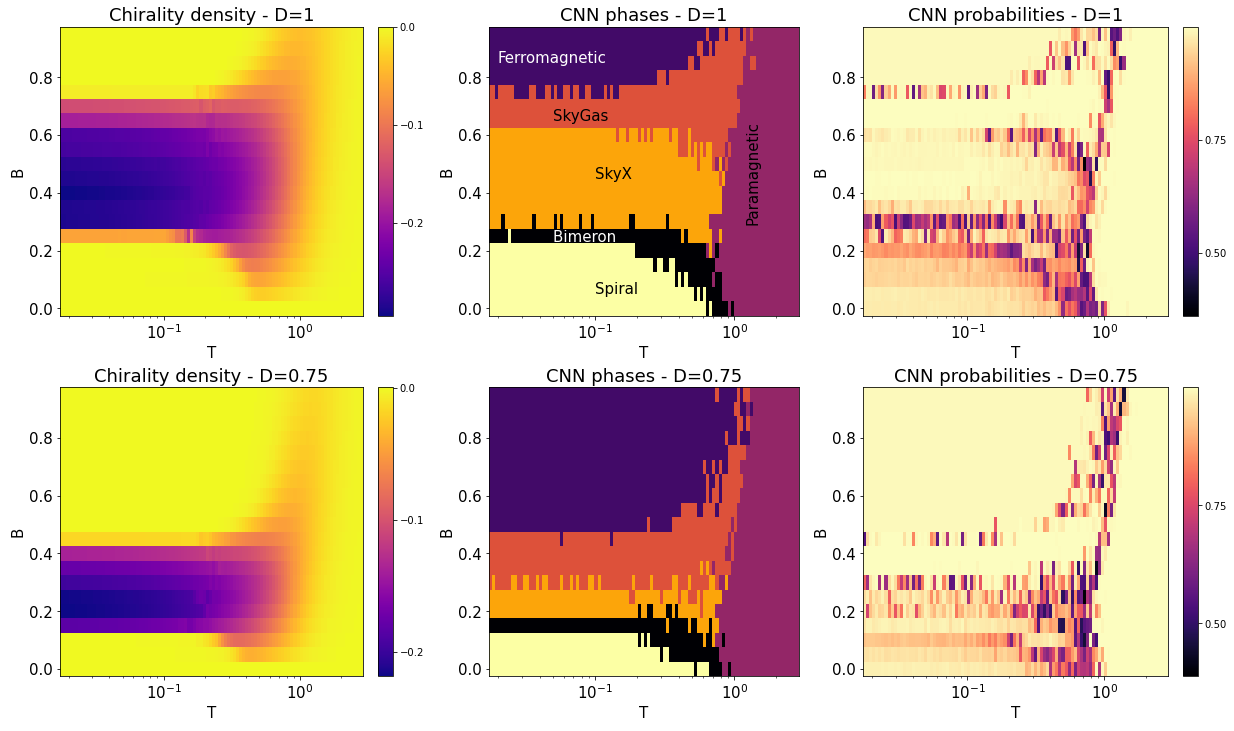}
\caption{\label{fig:pdCNN1} Phase diagrams for $D=1$ (top) and $D=0.75$ (bottom). First column: chirality density obtained from MC simulations. Middle column: phases predicted by the CNN. Right column: probabilies for the phases predicted by the CNN  }
\end{figure*}

To further explore the network, to see how to get the low temperature phases in the test dataset, we plot the probability of all six phases as a function of magnetic field, calculated for all low temperature configurations in the test dataset, $D=1.25$ and $D=0.75$, at the top row of Fig.~\ref{fig:probstest}. The bottom row shows the low temperature chirality as a function of magnetic field, calculated as the mean value and standard deviation over 30 independent realizations per field of the thermal average obtained from MC simulations. In the $D=1.25$ case, for the chosen range of $B$, it can clearly be seen that the CNN identifes the low $B$ configurations as spirals, and the higher $B$ snapshots as skyrmion crystal. At a certain point of $B$, the probability for the bimeron phase raises significantly, indicating a small Bm phase at low $T$. Inspection of the chirality agrees with this picture: it is zero at low fields, and it has a finite and constant value at higher $B$. Between these values,  intermediate points can be identified, matching the Bm prediction. For $D=0.75$, more phases can be seen in a smaller range of magnetic field. We may also identify intermediate regions, so that the selected phases as a function of the magnetic field are spiral, bimeron, skyrmion crystal, skyrmion gas and paramagnetic. No high probability can be seen for the paramagnetic phase for either value of $D$, which is consistent with the fact that we are inspecting low temperature configurations.

We now proceed to construct the complete $B-T$ phase diagram, applying the trained CNN to snapshots for all temperatures and magnetic fields . In Fig.~\ref{fig:pdCNN1} we plot the resulting phase diagrams for $D=1.0$, where only the lowest $T$ and the $B=0$ and $B=0.9$ high $T$ configurations where used for training and validation, and for $D=0.75$, which was not used to train the CNN. To compare the results, we plot the chirality density obtained from Monte Carlo simulations. We can see that there is a very good agreement between both phase diagrams and the $\chi$ density. Moreover, we can see that there is no ``phase mixing'': the CNN does not assign for example a spiral at high magnetic field, or an ordered phase at high temperature.

\begin{figure}[h!]
\includegraphics[width=0.99\columnwidth]{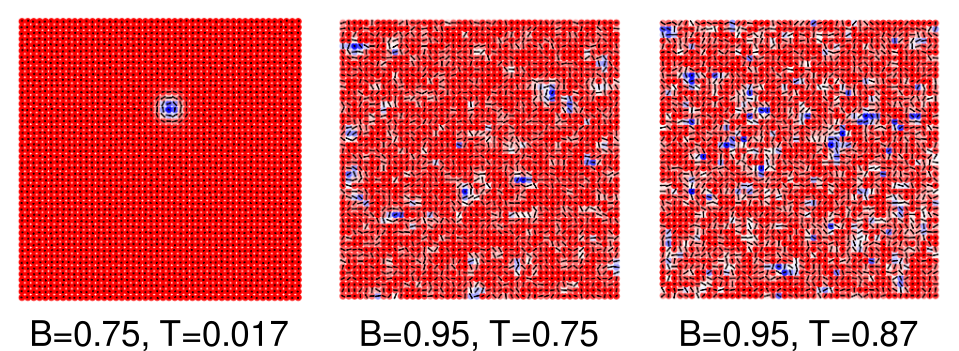}
\caption{\label{fig:exskg} Examples of snapshots classified as skyrmion gas, $D=1$ Left: one single skyrmion at the lowest temperature, $T=0.017$, for $B=0.75$ Center: skyrmion gas intermediate phase induced by temperature for $T=0.75$, $B=0.95$ Right: snapshot of the configuration at $T=1.0$, $B=0.95$ }
\end{figure}

 Although the CNN seems to overestimate slightly some intermediate regions, for example the SkG phase at high fields, or the Bm phase at very low fields, it can also correctly identify what may be considered ``tricky'' configurations, such as single skyrmions in a ferromagnetic background. We plot in Fig.~\ref{fig:exskg} three snapshots at high field that the CNN labels as SkG: one is a single well formed skyrmion at low $T$ (left) , and the other two are at higher temperatures: in one case, skyrmion like structures may still be identified (center), in the other one, thermal fluctuations have destroyed the characteristic skyrmion like structures.

In order to continue analyzing the output of the CNN, in the right column of Fig.~\ref{fig:pdCNN1} we plot the phase diagram of the maximum probability assigned to the chosen class. It can be seen that the CNN classifies with high probability in the middle of the different phases, but it gets lower in the interphases, both with magnetic field and temperature. In this sense, the probability gives us valuable physical information, specially in regions whith stronger thermal fluctuations. 

\begin{figure}[h!]
\includegraphics[width=1.0\columnwidth]{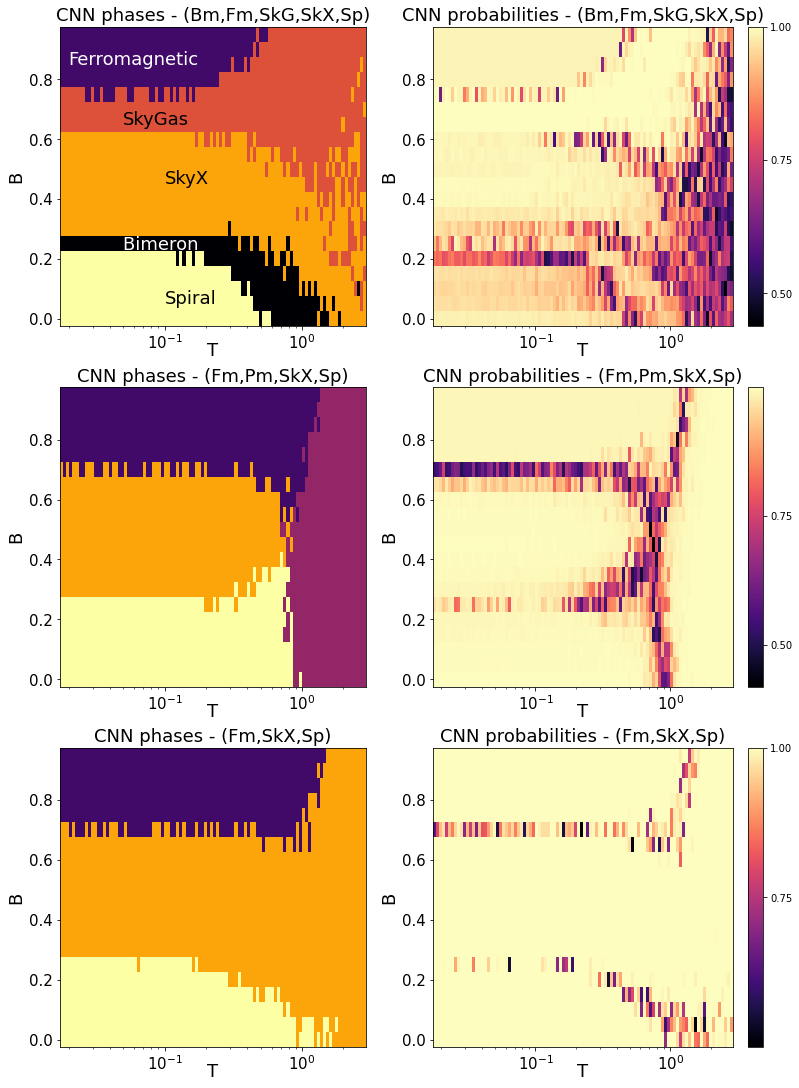}
\caption{\label{fig:pdCNN2} Phase diagrams (left) and probabilities (right) assigned to each phase obtained with the trained CNN for $D=1$ and three different subsets of the data set: five phases without the paramagnetic case (Bm, Fm, SkG, SkX, Sp), three phases with the paramagnetic phase - no intermediate phases (Fm, SkX, Sp, Pm), three phases without intermediate or paramagnetic cases (Fm, SkX, Sp)  }
\end{figure}

We may then wonder two things: whether it is necessary to include a paramagnetic phase, and what would change if we only classified with the three well defined phases (Fm, SkX, Sp). As was stated in previous works \cite{MLSky,MLSky2}, fluctuations in the probability may gives us an insight on whether to expect intermediate or different phases. We thus proceed as we did before for the six phase dataset, constructing three different subsets: five low $T$ phases (Bm, Fm, SkG, SkX, Sp), three low $T$ phases and the paramagnetic one (Fm, Pm, SkX, Sp) and only three low $T$ phases (Fm, SkX, Sp). For each case, a different CNN model was trained, but we mantained the architecture of the network. As a comment, excluding the intermediate Bp and SkG phases enhances the accuracy: for the last two data subsets the accuracy is 100$\%$ for the training, validation and test sets. We plot the phase diagrams and the corresponding probabilities for $D=1$ in Fig.~\ref{fig:pdCNN2}.  

First, we see that including the paramagnetic phase in the dataset is helpful to define the temperature limits of the low $T$ configurations. In the case where we included intermediate phases but excluded the paramagnetic class (Bm, Fm, SkG, SkX, Sp) Fig.~\ref{fig:pdCNN2}, top, we see that at low magnetic fields the assigned probabilities drop at higher temperature, suggesting that the classification does not work there (and thus, that a different phase may have to be considered), although this may  not said looking at higher magnetic fields. Then, if we only take the three phases that would be stable in the $T \rightarrow 0$ limit (spirals, skyrmions crystals and ferromagnetic), we see that the inclusion of the paramagnetic phase (Fm, Pm, SkG, SkX, Sp) does not only help to stablish the temperature limits of these low $T$ phases, but it also favors the definition of possible intermediate phases, given by the areas where the probabilities are lower, both with temperature and magnetic field (Fig.~\ref{fig:pdCNN2}, middle panel). This is clearly in contrast with taking only the three low $T$ phases (Fm, SkX, Sp), bottom panel of Fig.~\ref{fig:pdCNN2}. In that case,  the probabilities drop in a narrow region in magnetic field, and clearly do not  drop at higher temperatures, showing that the CNN underestimates the intermediate phases, specially the bimerons at low magnetic fields. Nonetheless, this comparison allows us to propose  that training a CNN with well defined phases, such as (Fm, Pm, SkX, Sp), and applying it to a whole range of temperature and magnetic field to unknown configurations also serves as a tool to identify regions where there might be possible new phases.

\subsection{Comparison with other classification methods}

\begin{figure}[t]
\includegraphics[width=0.95\columnwidth]{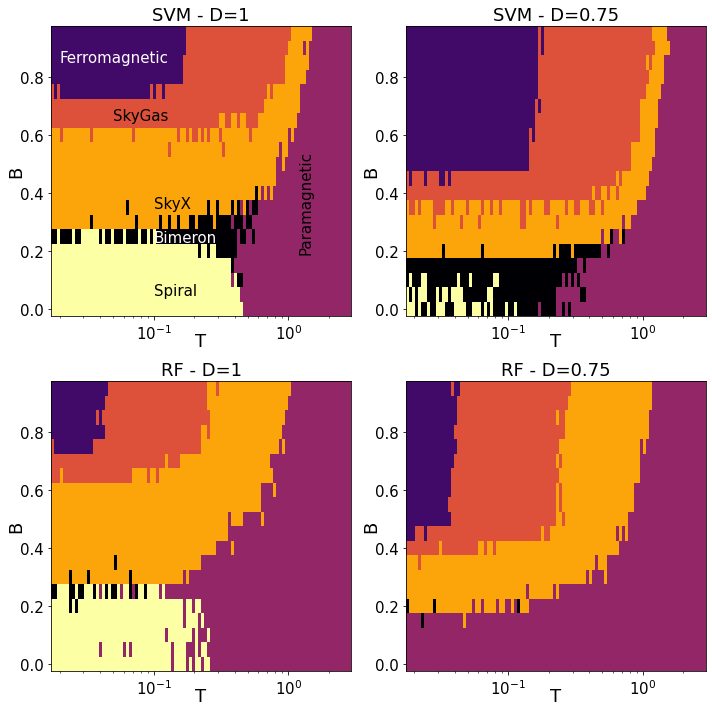}
\caption{\label{fig:svmrf} Phase diagrams obtained with SVM (top) and RF (bottom), for $D=1$ (left column) and $D=0.75$ (right column)  }
\end{figure}

In this subsection, we resort to other less complex ML classification techniques, to compare with the CNN.  First, we choose the Support Vector Machine (SVM) \cite{Bishop,Geron}. Using the Scikit Learn package, we proceed as before, with the 6 class dataset (Bm, Fm, Pm, SkG, SkX, Sp). We use cross validation to optimize the regularization  parameter 'C' and  the kernel coefficient 'gamma', choosing a radial basis function as kernel.  The resulting SVM has a 99.9$\%$ accuracy in the training set, and 95$\%$ in the validation set. However, the accuracy drops when applying it to the test set, where it goes to 78$\%$. Inspection of the CM shows that the most notable problem lies in the spiral class: more than half of spiral the configurations are classified as  paramagnetic. Given that the test set is formed by other values of $D$ not used in training, we may say that the SVM does not generalize as well as the CNN. We compare the complete $B-T$ phase diagrams for $D=1$ and $D=0.75$ in Fig.~\ref{fig:svmrf}, top row, where we confirm this last statement. For $D=1$, which was used for training,  the phase diagram is schematically correct at low temperatures. As the temperature increases, it is still consistent with the known phase diagram at low fields, but fails at higher fields, specially where there is an intermediate SkX high $T$ and high $B$ phase. This is even worse for $D=0.75$, where the classification is  problematic even at low temperature. Then,  the SVM may be a good tool to explore a first version of a phase diagram, provided that $D$ (basically, skyrmion and spiral size) is fixed.

Secondly, we train a Random Forest (RF) \cite{Bishop,Geron}, which is a collection of Decision Trees (DT) classifiers. As before, using cross validation we optimized the number of DTs and the maximun depth of each DT. As with the SVM case, the accuracy in the training set is high, 100$\%$, but it drops to $89\%$ in the validation set. It gets worse when changing the size of the textures: the accuracy is $43\%$ in the test set, showing that the RF is not a good tool to extend the classification to other configurations. We illustrate this by constructing the phase diagrams, shown in the bottom row of Fig.~\ref{fig:svmrf}. For $D=1$, we see that the RF reproduces the behavior at low $T$, consistent with the accuracy score in the training set, but it does not generalize well when applying it to higher temperature configurations. Clearly, inspecting the $D=0.75$ case, we see that the RF practically does not detect intermediate phases, and fails dramatically at low magnetic fields, where it assigns a paramagnetic phase at low temperature. This further supports the claim that this technique does not generalize well to values of $D$ that have not been used in training.

\section{Conclusions} 

Our main goal in this work was to present a machine learning approach   to classify different types of topological phases, including skyrmions and bimerons, to construct a complete detailed phase diagram,  and  that it would be able to generalize to different skyrmion sizes. To this end, we chose a  Convolutional Neural Network and we constructed the dataset from Monte Carlo simulations, enhanced for intermediate phases (bimerons and skyrmion gas). We trained and validated the CNN using only the lowest temperature snapshots, excluding other measurements or parameters from the Hamiltonian used to generate this configurations. In this way, our resulting trained CNN would be applicable to other configurations or images, provided they are formatted as the input data. The training and validation sets where chosen for  fixed values of the DM interaction, i.e. three characteristic skyrmion and spiral sizes. To test whether the CNN can generalize to other skyrmion sizes, we chose as the test set two different DM values. Since we are interested in building a complete phase diagram, we included high-temperature snapshots in the paramagnetic phase.

The resulting CNN model was then applied to configurations for a wide range of temperature and magnetic field, for different values of $D$, to construct the complete and detailed $B-T$ phase diagram. Comparing with the chirality density obtained from Monte Carlo simulations, we find a remarkable agreement, even in the intermediate phases.  This shows that   CNNs are  a powerful tool not only to construct a complete phase diagram, but to extend this type of work to other models. Given that the only input for the CNN is the spin configuration, the resulting trained CNN models are applicable to snapshots obtained with simulations for other Hamiltonians, or even real-space experimental images. 

We also discuss the importance of including paramagnetic and intermediate phases, constructing the phase diagram for different subsets of the data. We see that, as a first approach, taking only the three well ordered low temperature phases may help to determine possible intermediate phases, although this fails at higher temperature unless the paramagnetic phase is included.

Finally, we used the same approach choosing two other machine learning classification techniques: Support Vector Machine and Random Forest. In both cases, there is a very high accuracy in the training set, but it drops significantly when applying the resulting algorithm to snapshots with different skyrmion and spiral sizes, specially in the RF case. This shows that these techniques, very powerful for other tasks, do not generalize as well as the CNNs. 

We hope this work contributes to the ongoing work resorting to machine learning techniques in condensed matter. In the future, we expect to continue this type of work in other non-trivial topological phases such as antiferromagnetic skyrmions \cite{Rosales2015,Villalba,Zuko1,Zuko2}, antiskyrmions \cite{Muk1,Muk2}, chiral phases \cite{Jaubert2017,AlbaPujol} and spin liquids \cite{Balents,Bilayer,Jaubert2016}.

\section*{Acknowledgments} 
The authors thank P. Pujol and L. Jaubert for helpful discussions. F. A. G. A. especially thanks R. F. Díaz, M. Szewc, L. A. Nieto  and everyone in the ML course at UNSAM for important input and for generating an amazing working environment. 
This work was partially supported by CONICET (PIP
2021-11220200101480CO), and SECyT UNLP
PI+D X792, X788 and X893. F. A. G. A. acknowledges support from PICT 2018-02968.


\section*{Appendix}

Here we describe in detail de architecture of the CNN network, which we mantained when analyzing all the subsets of data, with the exception of the last layer, where the number of units matches the number of classes (phases).

 The CNN was constructed with four Convolutional Layers with 32 filters of size 3, no padding and activation function ReLu. Each of these layers was followed by a Max Pooling layer of pool size 2 and a Dropout layer with dropout value of 0.25. Then, the system was flattened and followed by a Dense layer of 128 nodes and activation function ReLu, a Dropout layer with dropout 0.25, and finally a Dense output layer with activation function softmax. 
 
 We chose cross entropy  as the error,   Scalar  Gradient Descent as the optimizer  and a learning rate 0.05. The CNN was trained in 200 epochs, with batch size 32. The CNN was implemented in TensorFlow \cite{TF}.



\clearpage
\onecolumngrid

\end{document}